\documentclass[%
reprint,
superscriptaddress,
amsmath,amssymb,
aps,
prl,
]{revtex4-1}

\usepackage{amsmath}
\usepackage{booktabs}
\usepackage{graphicx}
\usepackage{epstopdf}
\usepackage{subfigure}
\usepackage{color}
\usepackage{float}
\usepackage{comment}
\usepackage{wrapfig}
\usepackage{lineno}
\usepackage{bm}
\usepackage{cases}
\usepackage{amssymb}
\usepackage{slashed}
\usepackage{setspace}
\usepackage{mathrsfs}
\usepackage{hyperref}
\hypersetup{colorlinks=true,linkcolor=blue,anchorcolor=blue,citecolor=blue,urlcolor=blue}
\usepackage{soul}

\begin{document}

\newcommand*{\affaddr}[1]{#1}
\newcommand*{\affmark}[1][*]{\textsuperscript{#1}}

\title{Extended Time-Dependent Density Functional Theory for Multi-Body Densities}

\author{Jiong-Hang Liang}
\affiliation{Key Laboratory for Laser Plasmas and School of Physics and Astronomy, and Collaborative Innovation Center of IFSA, Shanghai Jiao Tong University, Shanghai, 200240, China.}
\author{Tian-Xing Hu}
\affiliation{Key Laboratory for Laser Plasmas and School of Physics and Astronomy, and Collaborative Innovation Center of IFSA, Shanghai Jiao Tong University, Shanghai, 200240, China.}
\author{D. Wu}
\email{dwu.phys@sjtu.edu.cn}
\affiliation{Key Laboratory for Laser Plasmas and School of Physics and Astronomy, and Collaborative Innovation Center of IFSA, Shanghai Jiao Tong University, Shanghai, 200240, China.}
\author{Zheng-Mao Sheng}
\email{zmsheng@zju.edu.cn}
\affiliation{Institute for Fusion Theory and Simulation, Department of Physics, Zhejiang University, Hangzhou 310027, China.}
\author{J. Zhang}
\affiliation{Key Laboratory for Laser Plasmas and School of Physics and Astronomy, and Collaborative Innovation Center of IFSA, Shanghai Jiao Tong University, Shanghai, 200240, China.}
\affiliation{Institute of Physics, Chinese Academy of Sciences, Beijing 100190, China.}

\date{\today}

\begin{abstract}
Time-dependent density functional theory (TDDFT) is widely used for understanding and predicting properties and behaviors of matter. As one of the fundamental theorems in TDDFT, van Leeuwen's theorem [Phys. Rev. Lett. 82, 3863 (1999)] guarantees how to construct a unique potential with the same one-body density evolution. Here we extend van Leeuwen's theorem by exploring truncation criteria in BBGKY-hierarchy. Our generalized theorem demonstrates the existence of a unique non-local potential to accurately reconstruct the multi-body density evolution in binary interacting systems. Under non-stringent conditions, truncation of the BBGKY-hierarchy equations aligns with the behavior of multi-body density evolution, and maintains consistency in the reduced equations. As one of applications within the extended TDDFT supported by our theorem, multiple excitation energy can be typically solved as the eigenvalue of a generalized Casida's equation. The extended TDDFT provides an accurate and first-principle framework capable of describing the kinetic processes of correlated system, including strongly coupled particle transport, multiple excitation and ionization processes.
\end{abstract}

\maketitle

Time-dependent density functional theory (TDDFT) is a powerful theoretical and computational tool, widely utilized for understanding and predicting diverse properties and behaviors of matter. It enables the exploration of excitation energy spectra in solids and complex molecular systems \cite{ullrich2011time, PhysRevLett.130.118401} and providing valuable insights into the unique characteristics and properties of warm dense matter \cite{PhysRevLett.116.115004}.

Within TDDFT, the Adiabatic Local Density Approximation (ALDA) effectively examines single-excitation processes by relying on properties like the linear response phenomenon. One successful application is the calculation of single excitation energies through Casida's equation \cite{casida1995time}, which serves as a notable example.

Despite ALDA-TDDFT's efficacy in addressing single-excitation phenomena, it encounters limitations when confronted with various physical scenarios, including ion-photoelectron entanglement control \cite{ishikawa2023control}, strong-field ionization \cite{lappas1998electron}, electron-ion scattering \cite{Appel_2010}, two-body dissipation in nuclear fusion reactions \cite{wen2018two}, as well as double-excitation and multiple-excitation processes \cite{elliott2011perspectives, rajam2009phase, levine2006conical, love1972multiple}. The intricacies of these scenarios demand theoretical frameworks beyond ALDA-TDDFT.

The challenges hindering TDDFT's capability to surpass ALDA stem from increasing perturbations leading to nonlinear effects, which impact system eigenstates and limit the viability of the adiabatic approximation. Additionally, the relationship between spatial and temporal localization within xc-potential functionals \cite{PhysRevLett.73.2244} indicates the need for multidimensional strategies, going beyond memory effect corrections \cite{PhysRevLett.55.2850}, to tackle the ultranonlocality problem \cite{ullrich2011time}.

Moreover, crucial properties and observables like correlation entropy and double-excitation probability rely on the two-body evolution behavior, posing challenges due to the idempotence of the KS 1-RDM (Reduced Density Matrix). Efforts to reconstruct the one-body density using delocalized orbitals in a two-electron system, as shown in Ref.\ \cite{rajam2009phase}, may introduce spurious oscillations for properties, like momentum density, not directly associated with the one-body density. In scenarios characterized by strong nonlinearities, such as electron-ion scattering and atoms in intense laser fields \cite{Appel_2010}, the absence of time-dependent natural occupation numbers (ONs) within the adiabatic approximation restricts accurate descriptions, which involves the one-body density functional representation problem of correlation entropy. 

These limitations have spurred the exploration of alternative methodologies. One promising approach involves augmenting the number of configurations beyond the confines of spatial-local one-body density. For instance, Time-Dependent Current Density Functional Theory (TDCDFT) confronts the ultranonlocality problem by establishing the relationship between current density and vector potential \cite{PhysRevB.70.201102, PhysRevLett.79.4878}. Unlike the application of TDDFT to periodic systems, TDCDFT operates unencumbered by the restrictions imposed by distinct boundary conditions \cite{PhysRevLett.77.2037}. Notwithstanding these advancements, complexity persist in representing observables such as double excitation within the $[n,j]$ functional in TDDFT. Meanwhile, Time-Dependent Density Matrix Functional Theory (TDDMFT) preserves the characterization of the 2-RDM, facilitating the delineation of ONs evolution \cite{PhysRevA.75.012506}. However, its efficay is contingent upon the selection of a 1-RDM functional. Adiabatic approximations, rooted in ground-state Density Matrix Functional Theory (DMFT), are insufficient to alter ONs without additional energy-minimizing procedures \cite{appel2010time, PhysRevA.81.042519}. Another method proposed, semiclassical TDDMFT, is anticipated to induce the time-dependent ONs, albeit through a purely classical correlation treatment \cite{PhysRevLett.105.113002}. Furthermore, time-dependent pair density functional theory (TD-PDFT) \cite{nagy2018time}, the time-dependent analogy to pair density functional theory (PDFT) \cite{ziesche1994pair}, is in its infancy. Accurately reconstructing the density evolution of Coulomb systems using a non-local potential description in TD-PDFT remains a challenge. This quandary is tied to the N-representability problem, delineating the thresholds at which the time-dependent 2-RDM can depict observable phenomena in an N-particle system \cite{PhysRevA.17.1257}.

In this Letter, to address the limitations of TDDFT in describing multi-body density evolution, we extend the van Leeuwen's theorem \cite{PhysRevLett.82.3863} to encompass the BBGKY-hierarchy equations \cite{bogoliubov1962studies}. We prove that for any order of the BBGKY equation, given the initial condition, there exists a unique non-local potential capable of reconstructing the multi-body density evolution within the binary interacting system. Consequently, under non-stringent conditions, the truncation of the BBGKY-hierarchy equations can be based on the behavior of the multi-body density evolution of interest. The reduced equations maintain consistency. Furthermore, we analyze linear response to demonstrate that the multi-particle excitation energy can be expressed as the eigenvalue of a generalized Casida's equation.

We start from a Hamiltonian $\hat{H}_{1...N}$ of a finite spinless particle system with binary interaction, $k$-body potential and $(k+1)$-body potential
\begin{equation}
    \begin{split}
    \hat{H}_{1...N}=\sum_{i=1}^N \left(\frac{\hat{\mathbf{p}}_{i}^2}{2m}+\hat{U}_{i}^{ext}(t)\right)+\frac{1}{2}\sum^{N}_{l\neq i}\hat{V}_{i,l}&\\
    +\frac{1}{k!}\sum_{i_1}^N...\sum_{i_k}^{N-k+1} \hat{W}_{i_1...i_k}\left(t\right)&\\
    +\frac{1}{\left(k+1\right)!}\sum_{i_1}^N...\sum_{i_{k+1}}^{N-k}\hat{Q}_{i_1...i_{k+1}}(t)&.
    \end{split}
    \label{equ14}
\end{equation}
By setting $\hat{Q}_{i_1...i_{k+1}}(t)=0$ and $\hat{W}_{i_1...i_k}\left(t\right)=0$, we may characterize a N-particle system with exclusively binary interactions. The system evolution behavior satisfies von Neumann equation
\begin{equation}
    i\partial_t\hat{\rho}_{1...N}\hat{\Lambda }^\pm_{1...N}=\left[\hat{H}_{1...N},\hat{\rho}_{1...N}\right]\hat{\Lambda }^\pm_{1...N}.
    \label{equ2}
\end{equation}
with the initial condition $\hat{\rho}_{1...N}\left(t_0\right)=\hat{\rho}^0_{1...N}$. The (anti-)symmetrization opertator $\hat{\Lambda }^\pm_{1...N}$ acts on the state to construct a (anti-)symmetric state \cite{boercker1979degenerate, bonitz2016quantum}. The $k$-body potential here should be a symmetric operator, $\hat{W}_{i_1...i_m...i_n...i_{k}}\left(t\right)=\hat{W}_{i_1...i_n...i_m...i_{k}}\left(t\right)$, due to the properties of (anti-)symmetric operator $\hat{\Lambda}_{1...N}$. And the same with $(k+1)$-body potential. Then, We have the equation of motion (EOM) of the $k$-RDM with the $(k+1)$-order collision term
\begin{equation}
    \begin{split}
    i\partial_t\hat{\rho}_{1...k}(t)&\hat{\Lambda }^\pm_{1...k}\\&=\left[\hat{H}_{1...k}(t),\hat{\rho}_{1...k}(t)\right]\hat{\Lambda }^\pm_{1...k}-\mathscr{L}_c(t).
    \end{split}
    \label{equ20}
\end{equation}
where the correlation term is
\begin{equation}
    \begin{split}
    \mathscr{L}_c(t)=&\frac{1}{\left(k+1\right)!}\sum_{i_1}^{k+1}...\sum_{i_{k+1}}^{1}\\
    &\quad\mathrm{Tr}_{k+1}\left[\hat{O}_{i_1...i_{k+1}}(t),\hat{\rho}_{1...k+1}(t)\right]\hat{\Lambda }^\pm_{1...k+1}.
    \end{split}
\end{equation}
Here the new $(k+1)$-body potential $\hat{O}_{i_1...i_{k+1}}(t)$ is an arbitrary symmetric operator. We may define the $(k+1)$-body potential in Hamiltonian (\ref{equ14}) by
\begin{equation}
    \begin{split}
\hat{Q}_{i_1...i_{k+1}}(t)=\hat{O}_{i_1...i_{k+1}}(t)-\sum_{l=1}^{k}\hat{V}_{i_l,i_{k+1}}&\\
-\sum_{l=1}^{k}\hat{W}_{i_1...i_{l-1},i_{l+1}...i_{k+1}}\left(t\right)&.
    \end{split}
\end{equation}
The $k$-body Hamiltonian in Eq.\ (\ref{equ20}) is
\begin{equation}
    \begin{split}
    \hat{H}_{1...k}(t)=\sum_{i=1}^k \left(\frac{\hat{\mathbf{p}}_{i}^2}{2m}+\hat{U}_{i}^{ext}(t)\right)+\frac{1}{2}\sum^{k}_{l\neq i}\hat{V}_{i,l}&\\
    +\frac{1}{k!}\sum_{i_1}^k...\sum_{i_k}^{1} \hat{W}_{i_1...i_k}(t)&.
    \end{split}
    \label{equ21}
\end{equation}
Thus we have the EOM of the second time-derivative of $k$-body density by calculating the zeroth and first moment of $k$-RDM \cite{[{See the Supplemental Material at }][{ for the derivation of the BBGKY-hierarchy for the modified Hamiltonian and the moment equations.}]section1}.
\begin{equation}
    \begin{split}
        \partial_t^2 n_{1...k}(t)-\sum_{i=1}^k\mathbf{\partial}_{i}\cdot[n_{1...k}(t)\mathbf{\partial}_{i}(\mathscr{U}_{i}(t)+\frac{1}{2}\sum_{l\neq i}\mathscr{V}_{i,l})]&\\
        =\mathbf{\partial}_{\tilde{\mathbf{R}}_k}\cdot\left[n_{1...k}(t)\mathbf{\partial}_{\tilde{\mathbf{R}}_k}\left(\mathscr{W}_{1...k}(t)\right)\right]+\mathscr{Q}_{1...k}(t)&,
    \end{split}
    \label{equ22}
\end{equation}
where the parameters $\tilde{\mathbf{R}}_k\equiv \left(\mathbf{R}_1,...,\mathbf{R}_{k}\right)$ is the abbreviation of the multidimensional spatial coordinates $\mathbf{R}_k$ and we set $\mathbf{\partial}_{\tilde{\mathbf{R}}_k}=\sum_{i}\mathbf{\partial}_i=\sum_{i}\mathbf{\nabla}_{\mathbf{R}_i}$ for convenience.  The number density $n_{1...k}(t)$ is the expectation value of $k$-RDM
\begin{equation}
    n_{1...k}(t)=\langle \tilde{\mathbf{R}}_{k}|\hat{\rho}_{1...k}\left(t\right)\hat{\Lambda }^\pm_{1...k}|\tilde{\mathbf{R}}_{k}\rangle,
    \label{equ8}
\end{equation}
and $\mathscr{Q}_{1...k}(t)$ is the sum of the kinetic term and drag force term
\begin{equation}
    \begin{split}
    \mathscr{Q}_{1...k}(t)=\sum_{i,l}^k\mathbf{\partial}_{i}\mathbf{\partial}_{l} T_{1...k}^{\left(il\right)}(t)-\mathbf{\partial}_{\tilde{\mathbf{R}}_{k}}\cdot\int \mathrm{d}\mathbf{R}_{k+1}&\\
    \times n_{1...k+1}(t)\mathbf{\partial}_{\tilde{\mathbf{R}}_{k+1}}\mathscr{O}_{1...k+1}(t)&.
    \end{split}
    \label{equ23}
\end{equation}
Here $\mathscr{W}_{1...k}(t)$ and $\mathscr{O}_{1...k+1}(t)$ is respectively the expectation values of the general $k$-body potential and ($k+1$)-body potential, and the momentum-stress tensor $T_{k}^{\left(il\right)}$ can be writtern as
\begin{equation}
    T_{1...k}^{\left(il\right)}(t)= \langle \tilde{\mathbf{R}}_{k}|\left[\frac{\hat{\mathbf{p}}_i}{2},\left[\frac{\hat{\mathbf{p}}_l}{2},\hat{\rho}_{1...k}(t)\right]\right]\hat{\Lambda }^\pm_{1...k}|\tilde{\mathbf{R}}_{k}\rangle,
    \label{equ10}
\end{equation}

After that,  we consider a primed system with Hamiltonian
\begin{equation}
    \begin{split}
    \hat{H}'_{1...N}=\sum_{i=1}^N \left(\frac{\hat{\mathbf{p}}^{'2}}{2m}+\hat{U}_{i}^{ext}(t)\right)+\frac{1}{2}\sum^{N}_{l\neq i}\hat{V}_{i,l}&\\
    +\frac{1}{k!}\sum_{i_1}^N...\sum_{i_k}^{N-k+1} \hat{W}'_{i_1...i_k}\left(t\right)&\\
    +\frac{1}{\left(k+1\right)!}\sum_{i_1}^N...\sum_{i_{k+1}}^{N-k}\hat{Q}'_{i_1...i_{k+1}}(t)&,
    \end{split}
    \label{equ24}
\end{equation}
where we set the primed $(k+1)$-body potential $\hat{O}'_{i_1...i_{k+1}}(t)$ as above process. Then we have a similar equation as Eq.\ (\ref{equ22}) but primed 

    \begin{equation}
        \begin{split}
            \partial_t^2 n'_{1...k}(t)-\sum_{i=1}^k\mathbf{\partial}_{i}\cdot[n'_{1...k}(t)\mathbf{\partial}_{i}(\mathscr{U}_{i}(t)+\frac{1}{2}\sum_{l\neq i}\mathscr{V}_{i,l})]&\\
            =\mathbf{\partial}_{\tilde{\mathbf{R}}_k}\cdot\left[n'_{1...k}(t)\mathbf{\partial}_{\tilde{\mathbf{R}}_k}\left(\mathscr{W}'_{1...k}(t)\right)\right]+\mathscr{Q}_{1...k}'(t)&,
        \end{split}
        \label{equ25}
    \end{equation}

Assuming that the primed system has identical density evolution, $n'_{1...k}(t)=n_{1...k}(t)$, we may subtract Eqs.\ (\ref{equ22}) and (\ref{equ25}) to find
\begin{equation}
    \begin{split}
    &\mathbf{\partial}_{\tilde{\mathbf{R}}_k}\cdot\left[n_{1...k}(t)\mathbf{\partial}_{\tilde{\mathbf{R}}_k}\bar{\mathscr{W}}_{1...k}(t)\right]= \bar{\mathscr{Q}}_{1...k}(t),
    \end{split}
    \label{equ26}
\end{equation}
where $\tilde{\mathscr{W}}_{1...k}(t)=\mathscr{W}_{1...k}(t)-\mathscr{W}'_{1...k}(t)$ and $\bar{\mathscr{Q}}_{1...k}(t)=\mathscr{Q}_{1...k}'(t)-\mathscr{Q}_{1...k}(t)$. Now we will construct the non-local potential $\mathscr{W}'_{1...k}$ based on the Eq.\ (\ref{equ26}). Since Eq.\ (\ref{equ22}) is a second-order partial differential equation, the initial conditions should be statisfied as the equality of the initial density
\begin{equation}
    n_{1...k}\left(t_0\right)= n'_{1...k}\left(t_0\right),
    \label{equ27}
\end{equation}
and the initial time derivative of density
\begin{equation}
    \partial_t n_{1...k}(t)|_{t=t_0}=\partial_t n'_{1...k}(t)|_{t=t_0}.
    \label{equ28}
\end{equation}

The constraints of Eqs.\ (\ref{equ27}) and (\ref{equ28}) are satisfied when the initial conditions of the two systems, $\hat{\rho}^0_{1...N}$ and $\hat{\rho}^{'0}_{1...N}$, are equal, which is a harsher initial condition. Then we specify the boundary condition $\bar{\mathscr{Q}}_{1...k}(t)=0$ at infinity, which implies that we need to choose a gauge like an arbitrary time-dependent and spatial-independent function $c\left(t\right)$ for $\mathscr{W}'_{1...k}\left(t\right)$. Now Eq.\ (\ref{equ26}) is a standard sturm-liouville equation, and has a unique solution for $\tilde{\mathscr{W}}_{1...k}\left(t\right)$ if $n_{1...k}\left(t\right)$ and $\bar{\mathscr{Q}}_{1...k}(t)$ are given. The time variable in Eq.\ (\ref{equ22}) enters as a parameter only, and thus we can construct the $k$-body potential under the Taylor series
\begin{equation}
    \mathscr{W}'_{1...k}\left(t\right)=\sum^\infty_{j=0}\frac{1}{j!} \mathscr{W}_{1...k}^{'(j)}\left(t-t_0\right)^j,
    \label{equ29}
\end{equation}

Let us start from $t=t_0$, the Eq.\ (\ref{equ26}) becomes
\begin{equation}
    \mathbf{\partial}_{\tilde{\mathbf{R}}_k}\cdot\left[n_{1...k}\left(t_0\right)\mathbf{\partial}_{\tilde{\mathbf{R}}_k}\left(\bar{\mathscr{W}}_{1...k}^{(0)}\right)\right]=\bar{\mathscr{Q}}_{1...k}(t_0).
    \label{equ30}
\end{equation}
Since the density $n_{1...k}$ are known at all times and $\mathscr{W}_{1...k}=0$ for Coulomb-interaction system, $\bar{\mathscr{Q}}(t_0)$ can be calculated from the given initial condition $\hat{\rho}^0_{1...N}$ and $\hat{\rho}^{'0}_{1...N}$, by Eq.\ (\ref{equ23}).

Next, we take the first time derivative of Eq.\ (\ref{equ26}) at the initial time $t=t_0$
\begin{equation}
    \begin{split}
    \mathbf{\partial}_{\tilde{\mathbf{R}}_k}\cdot\left[n_{1...k}\left(t_0\right)\mathbf{\partial}_{\tilde{\mathbf{R}}_k}\bar{\mathscr{W}}_{1...k}^{'(1)}\right]+\partial_t\bar{\mathscr{Q}}_{1...k}(t)|_{t=t_0}&\\
    =\mathbf{\partial}_{\tilde{\mathbf{R}}_k}\cdot\left[\partial_t n_{1...k}\left(t\right)|_{t=t_0}\mathbf{\partial}_{\tilde{\mathbf{R}}_k}\bar{\mathscr{W}}_{1...k}^{(0)}\right]&.
    \end{split}
    \label{equ31}
\end{equation}
Two time-derivative terms in Eq.\ (\ref{equ31}) are related to the first time derivative of $k$-RDM, $\partial_t\hat{\rho}_{1...k}\hat{\Lambda }^\pm|_{t=t_0}$ and $\partial_t\hat{\rho}'_{1...k}\hat{\Lambda }^\pm|_{t=t_0}$, which can be determined by Eq.\ (\ref{equ20}), with the given initial condition
\begin{equation}
    \begin{split}
    i\partial_t\hat{\rho}'_{1...k}(t)&\hat{\Lambda }^\pm_{1...k}|_{t=t_0}\\
    =&\left[\hat{H}'_{1...k}(t),\hat{\rho}'_{1...k}(t)\right]\hat{\Lambda }^\pm_{1...k}|_{t=t_0}+\mathscr{L}_c'(t),
    \end{split}
    \label{equ32}
\end{equation}
where the Hamiltonian $\hat{H}'_{1...k}=\hat{T}'_{1...k}+\hat{U}'_{1...k}+\hat{V}'_{1...k}+\mathscr{W}^{'(0)}_{1...k}$ can be determined by the zero-order of the $k$-body potential.

The procedure can be repeated for the $j$-th order of the time derivatives of Eq.\ (\ref{equ26})
\begin{equation}
    \begin{split}
    \mathbf{\partial}_{\tilde{\mathbf{R}}_k}\cdot\left[n_{1...k}\left(t_0\right)\mathbf{\partial}_{\tilde{\mathbf{R}}_k}\bar{\mathscr{W}}_{1...k}^{'(j)}\right]+\partial_t^j \bar{\mathscr{Q}}_{1...k}(t)|_{t=t_0}&\\
    =\sum_{i=1}^{j}\mathbf{\partial}_{\tilde{\mathbf{R}}_k}\cdot\left[\partial^i_t n_{1...k}\left(t\right)|_{t=t_0}\mathbf{\partial}_{\tilde{\mathbf{R}}_k}\bar{\mathscr{W}}_{1...k}^{(j-i)}\right]&.
    \end{split}
    \label{equ33}
\end{equation}
The coefficients $\mathscr{W}_{1...k}^{'(j)}$ in Taylor expasion (\ref{equ29}) can be determined by the initial condition and the lower-order time derivatives of the nonlocal potential $\mathscr{W}_{1...k}^{'(j-i)}$ that have already been calculated. Thus, the constructive procedure determines the nonlocal potential completely.

Now we consider two special cases. Firstly, we take $\hat{Q}_{i_1...i_{k+1}}(t)=0$ and $\hat{W}_{i_1...i_k}\left(t\right)=0$, which means that we consider a many-particle system with exclusively binary interaction. Let $\hat{O}'_{i_1...i_{k+1}}(t)=\hat{O}_{i_1...i_{k+1}}(t)$, we conclude that for a given initial condition $\hat{\rho}^{'0}_{1...N} = \hat{\rho}^{0}_{1...N}$, with the same initial condition constraints (\ref{equ27}) and (\ref{equ28}), there is a unique potential $\hat{W}'_{i_1...i_k}\left(t\right)=\hat{W}_{i_1...i_k}\left(t\right)$ that yields the given multi-body density evolution $n_{1...k}\left(t\right)$. It is actually the generalized Runge-Gross theorem containing the $k$-body correlation effects. The theorem implies a one-to-one correspondence between the multi-body potential and the multi-body density.

Then we take $\hat{O}'_{i_1...i_{k+1}}(t)=0$. Assuming that $\hat{\rho}^{'0}_{1...N} = \hat{\rho}^{0}_{1...N}$ with the correct initial density and time derivative of the density, there is a unique potential $\hat{W}'_{i_1...i_k}\left(t\right)$ (including a time-dependent function $c(t)$) which produces the multi-density evolution $n_{1...k}\left(\tilde{\mathbf{R}}_k,t\right)$ at $t>t_0$. This special case also implies that we construct a Liuville-type equation to reproduce the multi-density evolution $n_{1...k}\left(\tilde{\mathbf{R}}_k,t\right)$ in the many-body system
\begin{equation}
    i\partial_t\hat{\rho }_{1...k}(t)-\left[\hat{H}^\mathrm{eff}_{1...k}\left(\left[n_{1...k}\right],t\right),\hat{\rho}_{1...k}(t)\right]=0.
    \label{equ34}
\end{equation}
or the corresponding multi-density functional Schr\"odinger-type equation \cite{[{See the Supplemental Material at }][{ for the derivation of the multi-density functional Schr\"odinger-type equation and the corresponding properties.}]section2}
\begin{equation}
    i\partial_t|\phi_{1...k}(t)\rangle-\hat{H}^\mathrm{eff}_{1...k}\left(\left[n_{1...k}\right],t\right)|\phi_{1...k}(t)\rangle=0.
    \label{equse}
\end{equation}
where the reconstructed $k$-body Hamiltonian is
\begin{equation}
    \begin{split}
    \hat{H}^\mathrm{eff}_{1...k}\left(\left[n_{1...k}\right],t\right)=\hat{T}_{1...k}+&\hat{U}_{1...k}^\mathrm{ext}(t)\\
    &\quad+\hat{U}_{1...k}^\mathrm{rc}\left(\left[n_{1...k}\right],t\right).
    \end{split}
    \label{equ35}
\end{equation}
Here the reconstructed potential $\hat{U}_{1...k}^\mathrm{rc}(\left[n_{1...k}\right],t)\equiv\frac{1}{2}\sum_{i,j}^{k}\hat{V}_{i,j}+\hat{W}'_{1...k}(\left[n_{1...k}\right],t)$ is determined by Eq.\ (\ref{equ29}) and $k$-body density is obtained by Eq.\ (\ref{equ8}). The overall (anti-)symmetrization operator $\hat{\Lambda}_{1...k}$ can be canceled. Eq.\ (\ref{equ34}) are consistent in arbitrary dimensions, the only difference being the description of correlation in the reconstructed potential. 

Let $k=1$, Eq.\ (\ref{equ34}) equals the 1-RDM evolution equation in the Kohn-Sham ensemble \cite{li1985kohn} solving the non-interacting $v$-representability problem as van Leeuwen theorem does. Let $k=2$, Eq.\ (\ref{equ34}) can be written as
\begin{equation}
    i\partial_t\hat{\rho}_{12}(t)-\left[\hat{H}^\mathrm{eff}_{12}\left(\left[n_{12}\right],t\right),\hat{\rho}_{12}(t)\right]=0.
    \label{equ38}
\end{equation}
The two-body Hamiltonian $\hat{H}^\mathrm{eff}_{12}$ and the 2-RDM $\hat{\rho}_{12}$ are both functionals of two-body density $n_{12}$. Because of the one-to-one correspondence between binary interaction and two-body density, we can find a unique 2-RDM $\hat{\rho}_{12}(t)$ to determine the physical observables $M(t)=\mathrm{Tr}_{12}\hat{M}(t)\hat{\rho}_{12}(t)\hat{\Lambda }^\pm_{12}$. The results demonstrate the N-representability of the 2-RDM of many-particle systems. 

Eq.\ (\ref{equ34}) describes the evolution of density matrices at any $k$-order, yet the actual system remains singular. Each observable fundamentally relies on the functional of $k$-body density, albeit presenting distinct levels of functional complexity. The balance between complexity and precision dictates the selection of the EOM.  

Here we give a further application of Eq.\ (\ref{equ34}) in linear response regime. The multiple excitation problem can be evaluated under weak external potential perturbation $\hat{H}^\mathrm{eff}_{1...k}(t)=\delta \hat{U}_{1...k}^\mathrm{ext}(t)+\hat{H}^\mathrm{eff0}_{1...k}$ where the assumption $\delta \hat{U}_{1...k}^\mathrm{ext}(t)\ll\hat{H}^\mathrm{eff0}_{1...k}$ allows adiabatic approximation. We find that the multiple excitation energy $\omega_{k}^q$ is the eigenvalue solution of the following generalized Casida's matrix equation \cite{[{See the Supplemental Material at }][{ for the derivation of the generalized Casida's matrix equation based on the extended TDDFT.}]section3}

\begin{equation}
    \begin{pmatrix}
        \mathbf{A}_{k} & \mathbf{B}_{k}\\
        \mathbf{B}_{k}^\dagger & \mathbf{C}_{k}
    \end{pmatrix}
    \begin{pmatrix}
        \mathbf{X}_{1...k}^p\\
        \mathbf{X}_{1...k}^h
    \end{pmatrix}
    =\omega_{k}^q
    \begin{pmatrix}
        \mathbf{I} & 0\\
        0 & -\mathbf{I}
    \end{pmatrix}
    \begin{pmatrix}
        \mathbf{X}_{1...k}^p\\
        \mathbf{X}_{1...k}^h
    \end{pmatrix}
    \label{equ50}
    \end{equation}
where the eigenvector $\mathbf{X}_{1...k}^{p(h)}$ is the (de-)excitation response function for the multiple excitation process. 

When $k=1$, Eq.\ (\ref{equ50}) precisely mirrors Casida's equation within linear-response TDDFT \cite{casida1995time}. For $k=2$, Eq.\ (\ref{equ50}) corresponds to the pp-RPA result \cite{PhysRevA.88.030501}, wherein the lifetime of excitons is substituted with the $k$-body interactions within the reconstructed potential. The generalized Casida's equation enables the determination of multiple excitation energy within the adiabatic approximation, avoiding the need to account for the intricate nonlinear processes in TDDFT.

Our findings can be summarized here: The advancement of TDDFT faces challenges due to the non-adiabatic correction required for local exchange-correlation functionals and the complexity in representing multi-body observables via one-body density functionals. As a result, we generalize the van Leeuwen theorem of time-dependent density functional theory to encompass the BBGKY scheme. This extension reveals that, given the initial conditions, a unique non-local potential can reconstruct the multi-body density evolution within a binary interacting system. Consequently, under non-stringent conditions, truncating the BBGKY-hierarchy equations based on the behavior of multi-body density evolution becomes feasible. The resulting equations exhibit consistent mathematical properties. Moreover, by employing linear response theory, we ascertain that the multiple excitation energy constitutes the eigenvalue solution of a generalized Casida's equation, solvable under adiabatic approximation.

This work has been supported by the National Natural Science Foundation of China (No. 12247137, 12305271 and No. 12075204), the Strategic Priority Research Program of Chinese Academy of Sciences (Grant No. XDA250050500), and the Shanghai Post-doctoral Excellence Program (No. 2022324). D. Wu thanks the sponsorship from Yangyang Development Fund. 
\bibliography{bibtex}

\end{document}